\documentclass[prd,twocolumn,showpacs,preprintnumbers,amsmath,amssymb]{revtex4}
\usepackage{graphicx,array,epsfig}
\usepackage{bm}
\usepackage{pstricks,pst-plot}
\usepackage{mathrsfs}
\usepackage{mathbbol}

\newcommand{\be}{\begin{equation}}
\newcommand{\ee}{\end{equation}}
\newcommand{\bea}{\begin{eqnarray}}
\newcommand{\eea}{\end{eqnarray}}
\newcommand{\bml}{\begin{mathletters} \baselineskip 10pt}
\newcommand{\eml}{\baselineskip 12pt \end{mathletters}}
\newcommand{\nn}{\nonumber}

\newcommand{\ha}{\frac{1}{2}}
\newcommand{\uMF}{u_{\mathrm{MF}}}
\newcommand{\cD}{\mathscr{D}}
\newcommand{\cL}{\mathscr{L}}

\newcommand{\la}{\lambda}

\newcommand{\ga}{\gamma}

\newcommand{\bra}{\langle}
\newcommand{\ket}{\rangle}

\newcommand{\n}{\mathbf n}

\newcommand{\Lx}{L_{\ssvc{x}}}
\newcommand{\Ly}{L_{\ssvc{y}}}
\newcommand{\Lz}{L_{\ssvc{z}}}

\newcommand{\ssx}{\ssvc{x}}
\newcommand{\ssy}{\ssvc{y}}
\newcommand{\ssz}{\ssvc{z}}

\newcommand{\sfrac}[2]{{\textstyle \frac{#1}{#2}}}

\newcommand{\fud}[2]{\frac{\delta #1}{\delta #2}}

\newcommand{\vc}[1]{\mbox{\boldmath$#1$}}
\newcommand{\svc}[1]{\mbox{\footnotesize\boldmath$#1$}}
\newcommand{\ssvc}[1]{\mbox{\scriptsize\boldmath$#1$}}

\newcommand{\pol}[1]{\mathfrak{P}_{\svc{#1}}}

\newcommand{\tr}{\mbox{tr}}

\newcommand{\refs}[1]{(\ref{#1})}
\newcommand{\Spl}{S_{\mathrm{PL}}}

\begin{document}


\title{Effective Actions for the SU(2) Confinement--Deconfinement Phase
Transition}

\author{Thomas Heinzl}
\email{theinzl@plymouth.ac.uk}
\affiliation{School of Mathematics and Statistics\\
University of Plymouth\\
Drake Circus, Plymouth, PL4 8AA, United Kingdom}

\author{Tobias Kaestner}
\email{kaestner@tpi.uni-jena.de}

\author{Andreas Wipf}
\email{wipf@tpi.uni-jena.de}


\affiliation{Theoretisch--Physikalisches Institut,
Friedrich--Schiller--Universit\"at Jena\\
Max--Wien--Platz 1, 07743 Jena, Germany}

\date{\today}

\begin{abstract}
We compare different Polyakov loop actions yielding effective
descriptions of finite--temperature $SU(2)$ Yang--Mills theory on
the lattice. The actions are motivated by a simultaneous
strong--coupling and character expansion obeying center symmetry
and include both Ising and Ginzburg--Landau type models. To keep
things simple we limit ourselves to nearest--neighbor
interactions. Some truncations involving the most relevant
characters are studied within a novel mean--field approximation.
Using inverse Monte--Carlo techniques based on exact \emph{geometrical}
Schwinger--Dyson equations we determine the effective couplings of
the Polyakov loop actions. Monte--Carlo simulations of these
actions reveal that the mean--field analysis is a fairly good
guide to the physics involved. Our Polyakov loop actions reproduce
standard Yang--Mills observables well up to limitations due to the
nearest--neighbor approximation.
\end{abstract}

\maketitle
\section{\label{secs1}Introduction}
The finite--temperature confinement--deconfinement phase transition in
$SU(2)$ Yang--Mills theory originally conjectured by Polyakov
\cite{polyakov:78} and Susskind \cite{susskind:79} is by now fairly well
established. The order parameter is the Polyakov loop,
\be\label{in1}
  \Lx \equiv \ha \, \tr \, \mathcal{T} \, \exp \Big( i
  \int_0^{\beta_T} \!\!d\tau  \, A^0 (\vc{x}, \tau)\Big) \; ,
\ee
a traced Wilson line that winds around the periodic Euclidean time
direction parameterized by $\tau$, $0 \le \tau \le \beta_T$ where
$\beta_T = 1/T$ is the inverse temperature. In the confined phase
the expectation value $\bra L \ket$ is zero, while it becomes
nonvanishing in the broken, deconfined phase. The Polyakov loop
transforms nontrivially under the center symmetry,
\be \label{in3}
  \Lx \to z \Lx \; , \quad z = \pm 1 \in \mathbb{Z}(2) \; .
\ee
Thus, above the critical temperature, $T = T_c$, this symmetry
becomes spontaneously broken. Lattice calculations have shown
beyond any doubt that the phase transition is second order with
the critical exponents being those of the 3$d$ Ising model
\cite{caselle:96,gliozzi:97,forcrand:01,pepe:02}. This is in
accordance with the Svetitsky--Yaffe conjecture
\cite{svetitsky:82,yaffe:82} which states in particular that
$SU(2)$ Yang--Mills theory (in 4$d$) is in the universality class
of a $\mathbb{Z}(2)$ spin model (in $3d$) with short--range
interactions. Hence, it should be possible to describe the
confinement--deconfinement transition by an effective theory
formulated solely in terms of the Polyakov loop $\Lx$. The most
general ansatz is given by a center--symmetric effective Polyakov
loop action (PLA) of the form \cite{svetitsky:86}
\begin{eqnarray}
  \Spl[L] &=& \sum_{\ssx} V[\Lx^2] +
  \sum_{\ssx\ssy}\Lx K^{(2)}_{\ssx\ssy}\Ly \nn\\
  &+& \sum_{\ssx\ssy\ssvc{u}\ssvc{v}} \Lx
  \Ly K^{(4)}_{\ssx \ssy \ssvc{u} \ssvc{v}}
  L_{\ssvc{u}} L_{\ssvc{v}} + \ldots\label{in5}
\end{eqnarray}
There is a potential term $V$, a power series in $\Lx^2$ living on
single lattice sites $\vc{x}$, plus hopping terms with kernels
$K^{(2n)}$ connecting more and more lattice sites $\vc{x},\vc{y},
\ldots$. It is important to note that $L$ is a dimensionless and
compact variable, $\Lx \in [-1,1]$. Thus, in principle, one is
confronted with a proliferation of possible operators that may
appear in the PLA \refs{in5}. One simplification arises due to the
Svetitsky--Yaffe conjecture implying that the kernels $K^{(2n)}$
should be short--ranged. Hence, upon expanding like for instance,
\be
  K^{(2)}_{\ssx \ssy} = \sum_{\ssvc{r}}
  \la_{r} \, \delta_{\ssy  , \ssx + \ssvc{r}} \; ,\label{in7}
\ee
one expects that the first few terms with small $r \equiv |\vc{x}
- \vc{y}|$ will dominate, i.e.~will have the largest couplings
$\la_r$. To check this expectation one needs a reliable method to
calculate the kernels $K^{(2n)}$ or, equivalently, the coupling
parameters inherent in them. A particularly suited approach is
introduced in the following section.

\section{\label{sec2}Inverse Monte--Carlo Method}
Inverse Monte--Carlo (IMC) is a numerical method to determine
effective actions \cite{falcioni:86,gonzalez-arroyo:87a}. The
latter are generically defined via
\be \label{imc1}
  \exp(-S_{\mathrm{eff}}[X]) \equiv \int \cD U \, \delta (X -
  X[U]) \exp(-S[U]) \; ,
\ee
where the $U$'s represent some `microscopic' degrees of freedom
and the $X$'s the effective `macroscopic' ones. In the spirit of
Wilson's renormalization group, these are obtained by integrating
out the $U$'s in favor of the $X$'s. It is important to
distinguish this `Wilsonian' notion of an effective action from
the 1--particle--irreducible (1PI) effective action which will
later on be employed as well (cf.~the recent remarks in
\cite{banks:04}).

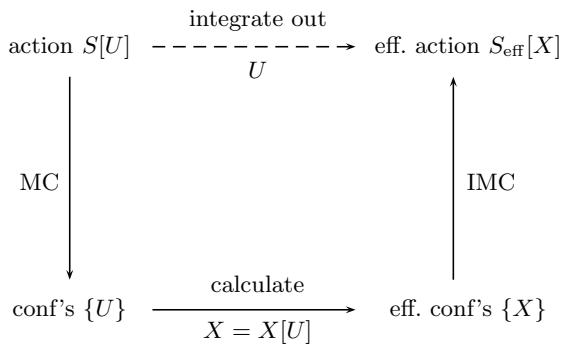
\begin{figure}[ht]
\psset{unit=1mm,linewidth=0.6pt}
\begin{pspicture}(-5,0)(75,45)
\rput(7,5){conf's $\{U\}$}
\rput(7,40){action $S[U]$}
\psline{->}(7,36)(7,9)
\rput(60,5){eff.\ conf's $\{X\}$}
\rput(60,40){eff.\ action $S_{\mathrm{eff}}[X]$}
\psline{->}(58,9)(58,36)
\psline{->}(18,5)(45,5)
\psline[linestyle=dashed]{->}(18,40)(45,40)
\rput(32,8.5){calculate}
\rput(32,2){$X=X[U]$}
\rput(32,43.5){integrate out}
\rput(32,37){$U$}
\rput(3,22){MC}
\rput(63,22){IMC}
\end{pspicture}
\caption{\label{fig:IMC}\textsl{Illustration of the IMC procedure.}}
\end{figure}

Of course, the problem with (\ref{imc1}) is to do the integration
nonperturbatively which in general is not possible. In this case,
one has to resort to choosing an ansatz like \refs{in5} as
dictated by symmetry and dimensional counting or to do the
integration numerically. The huge number of degrees of freedom
involved clearly suggests to  use Monte--Carlo (MC) methods.
However, this amounts to calculating expectation values rather
than integrals like \refs{imc1}. Hence, one needs a recipe to get
\emph{from expectation values to effective actions}. This is
exactly what IMC is supposed to do (see Fig.~\ref{fig:IMC}).

Our particular IMC method is based on the Schwinger--Dyson
equations that must hold in the effective theory once a particular
ansatz is chosen. In our case, the macroscopic degrees are given
by the Polyakov loops $\Lx$ distributed according to the PLA
$\refs{in5}$  which we rewrite as
\be \label{imc3}
  \Spl [L] \equiv \sum_{a=1}^{N_a} \la_a S_a [L] \; ,
\ee
with $\mathbb{Z}(2)$ symmetric operators $S_a$ and coupling
parameters $\la_a$ to be determined from the Schwinger--Dyson
equations. To derive these, we proceed as follows.

On a lattice with spacing $a$ and temporal extent $N_t$ (hence
temperature $T = 1/N_t a$) the Polyakov line is given by the product of
temporal links,
\be \label{imc5}
  \pol{x} \equiv \prod_{t=1}^{N_t} U_{\ssx,t;0} \in SU(2) \; .
\ee
This $SU(2)$ matrix may be diagonalized whereupon it can be
written as
\be
  \pol{x} = \left( \begin{array}{cc}
  \exp (i \theta_{\ssx}) & 0 \\
  0 & \exp (- i \theta_{\ssx})
                   \end{array}\right) \; , \nn
\ee
with $-\pi \le \theta_{\ssx} \le \pi$. This representation
immediately yields the trace (divided by two),
\be\label{imc7}
  \Lx = \sfrac{1}{2} \tr \pol{x} = \cos \theta_{\ssx} \; ,
\ee
which contains all the gauge invariant information contained in
the group variable $\pol{x}$. As an aside, we remark that this
peculiar feature will no longer be true for higher $SU(N)$ groups
\cite{meisinger:02a,dumitru:03}. In this more general case, traces
of $N-1$ different powers of $\pol{}$ are required.

With \refs{imc1} the action \refs{imc3} leads to
the partition function
\be \label{imc9}
  Z = \int \cD L \exp (- \Spl [L]) \; ,
\ee
where the integration is performed with the reduced Haar
measure $h$ of $SU(2)$,
\be \label{imc11}
  \cD L \equiv \prod_{\ssx}dh \, (\Lx) , \quad
  dh(u) = \frac{2}{\pi} \sqrt{1 - u^2} \, du \; .
\ee
Since $\Spl$ depends on the Polyakov loop only via the class
function $\Lx$ in \refs{imc7} we may use the left--right invariant
Haar measure $\cD \pol{}$ in \refs{imc9} instead of the reduced
Haar measure.  The enhanced  symmetry of the measure yields the
following \emph{geometrical} Schwinger-Dyson equations
\cite{dittmann:03},
\be \label{imc13}
  0 = \int \cD L \,  \exp (-\Spl) \left[ 3 \Lx G - (1 - \Lx^2)
  (G_{\ssx}^\prime - G
  S_{\mathrm{PL}, \ssx}^\prime) \right] \; .
\ee
Here, $G[L]$ represents some set of functions of the Polyakov loop
to be chosen appropriately (see below).  In addition, we have
defined the derivative $G_{\ssx}^\prime \equiv \partial G /
\partial \Lx$ and analogously for $\Spl$.

Switching to expectation value notation, \refs{imc13} can be rewritten
as a linear system for the couplings in \refs{imc3},
\be \label{imc15}
  \sum_a \bra (1 - \Lx^2) G S'_{a, \ssx}\ket \la_a = \bra (1 -
  \Lx^2) G'_{\ssx} \ket - 3 \bra \Lx G \ket \; .
\ee
The coefficients of this system are expectation values which are
calculated in the full Yang--Mills ensemble obtained by MC
simulation based on the $SU(2)$ Wilson action. Numerically, it is
of advantage to have more equations than unknown couplings
$\la_a$. This is achieved by choosing $G$ out of the following set
of local functions,
\be \label{imc17}
  G_{a,\ssy} \in \{ \partial S_a / \Ly \; , a = 1, \ldots, N_a \} \; ,
\ee
which represent the operators present in the equation of motion
for $\Ly$. For \textit{fixed} $\vc{y}$, \refs{imc15} then yields
as many equations as there are couplings, namely $N_a$. These
equations relate different two--point functions labeled by lattice
sites $\vc{x}$ and $\vc{y}$ of distance $r$. Additional relations
are obtained by letting the distance $r$ run through (half of) the
spatial lattice extent, $r = 1, \ldots, N_s /2$. Altogether, the
overdetermined system \refs{imc15} consists of $N_a \times  N_s
/2$ equations which are solved by least--square methods. For more
details the reader is referred to Appendix~B and
\cite{dittmann:03}.

\section{\label{sec3} Character expansion}
To find a reasonable choice of operators for the ansatz
(\ref{imc3}) we use the beautiful analytical results of Bill{\'o}
et al.\ \cite{billo:96}. These authors have evaluated the integral
(\ref{imc1}) for the case at hand ($S[U]$ being the Wilson action,
$X \equiv L$ the Polyakov loop) by combining the strong--coupling
with a character expansion. For the benefit of the reader we
briefly recapitulate their approach before we adopt it for our purposes.

Recall that a character is the trace of a group element in an
irreducible representation. If $j$ denotes the spin of an $SU(2)$
representation such that $p=2j$ is the length of the corresponding
Young tableau, then the associated character is
\be
\label{char1}
  \chi_p (U)= \tr_p (U) =\! \frac{\sin\big((p + 1) \theta
  \big)}{\sin \, \theta} \,,\;\; p=0,1,2,\ldots\;.
\ee
The characters can be entirely expressed as orthogonal polynomials
in the traced loop $L$, namely the Chebyshev polynomials of the
second kind \cite{abramowitz:64},
\be
  \chi_p (U) = \sum_{k=0}^{[p/2]} \frac{(-1)^k (p-k)!}{k!
  (p-2k)!} (2L)^{p-2k} \; .\nn
\ee
This representation manifestly shows that $\chi_p$ is a
polynomial in $L$ of order $p$. The first few characters are
\begin{eqnarray}
  \chi_0 = 1\quad&,&\quad
  \chi_1 = 2L \; ,  \nn \\
  \chi_2 = 4L^2 - 1&,&
  \chi_3 = 8L^3 - 4L \; .\label{char3}
\end{eqnarray}
To determine the PLA $\Spl$ (\ref{imc3}) in the strong coupling
limit of the underlying $SU(2)$ gauge theory Bill{\'o} et al.\
\cite{billo:96} allowed for different couplings in temporal and
spatial directions, denoted $\beta_t$ and $\beta_s$, respectively.
In terms of these couplings the original Wilson coupling becomes
\be
  \beta = \frac{4}{g^2} = \sqrt{\beta_t \beta_s} \; .\nn
\ee
The formula for the temperature,
\be
  T = \frac{1}{N_t} \sqrt{\frac{\beta_t}{\beta_s}} \; ,\nn
\ee
shows that the high--temperature limit (for $\beta_t$ fixed)
corresponds to $N_t$ or $\beta_s$ being small. An expansion in
terms of $\beta_s$ results in the PLA \cite{billo:96}
\be
\label{char5}
  \Spl \equiv  \sum_{\langle\ssx\ssy\rangle} \ln \left \{ 1 +
  \sum_{p = 1}^\infty \kappa_p \chi_p (\Lx) \chi_p (\Ly) \right\}+
\ldots \;
\ee
where all orders in $\beta_t$ have been summed up in terms of
coupling coefficients $\kappa_p = \kappa_p (\beta_t)$. The terms
not written explicitly contain higher orders in $\beta_s$ and
interactions of characters of plaquette type. The leading--order action
(\ref{char5}) involves only \emph{nearest--neighbor} (NN) interactions
with the couplings given explicitly by
\be
\label{char7}
  \kappa_p (\beta_t) = - \left[ \frac{I_{p+1}(\beta_t)}{I_1
(\beta_t)}\right]^{N_t} \; .
\ee
Asymptotically, for small $\beta_t$, this is
\be
\label{char9}
  \kappa_p =-c_p \beta_t^{\,pN_t}+O\big( \beta_t^{pN_t + 2}
  \big), \;\;\; c_p \equiv [2^p (p+1)!]^{-N_t}.
\ee
This concludes our brief discussion of \cite{billo:96}. In order
to make the operator (i.e.\ character) content of the action
(\ref{char5}) more explicit we expand the ln in \refs{char5} in
powers of $\beta_t$.
From the small--$\beta_t$ behavior \refs{char9}, we infer that a
product of $n$ $\kappa$'s behaves as
\be
  \kappa_{p_1} \ldots \kappa_{p_n} = O \big(\beta_t^{pN_t}\big) \; ,
  \quad p \equiv \sum_{i=1}^n p_i \; .\label{char11}
\ee
Thus we reshuffle the expansion of \refs{char5} such that, for
fixed $p$, we first sum over all partitions of the integer $p$,
then increase $p$ by one unit, sum again etc.\ up to some maximal
value, say $p=3$. In this way we obtain
\be \label{char13}
  \Spl = S^{(1)} + S^{(2)} + S^{(3)} + O\big(\beta_t^{4N_t}\big) ,
\ee
where $S^{(p)}$ is $O\big(\beta_t^{pN_t}\big)$. Accordingly, we
have a hierarchy of actions $S^{(p)}$ that become more and more
suppressed (for small $\beta_t$) as $p$ increases. We thus refer
to $S^{(1)}$ as being of leading order (LO), $S^{(2)}$ of
next--to--leading order (NLO) and so on. Abbreviating
$\chi_{p\ssx}=\chi_p(\Lx)$ the actions $S^{(p)}$ read explicitly
\begin{eqnarray*}
  S^{(1)}&=&\sum_{\langle\ssx\ssy\rangle}
\kappa_1 \chi_{1\ssx}\chi_{1\ssy} \; , \\
  S^{(2)}&=&\sum_{\langle\ssx\ssy\rangle}
\Big(\kappa_2 \chi_{2\ssx}\chi_{2\ssy}
  -\ha\kappa_1^2\chi_{1\ssx}^2\chi_{1\ssy}^2\Big) \; , \\
  S^{(3)}&=&
  \sum_{\langle\ssx\ssy\rangle}\Big( \kappa_3 \chi_{3\ssx}\chi_{3\ssy}
  -\kappa_1\kappa_2\chi_{1\ssx}\chi_{2\ssx}\chi_{1\ssy}\chi_{2\ssy}\\
  &&\qquad + \frac{1}{3}\kappa_1^3\chi_{1\ssx}^3\chi_{1\ssy}^3\Big) \; .
\end{eqnarray*}
The product of characters at the \emph{same site} may be
further reduced by
the $SU(2)$ `reduction formula',
\be
  \chi_{p_1} \chi_{p_2} = \chi_{p_1 + p_2} + \chi_{p_1 + p_2 -2} +
  \ldots + \chi_{|p_1 - p_2|} \; .\nn
\ee
Note that our conventions are such that the subscripts get reduced by
two units from left to right.
Using this formula in the above expressions we end up
with
\begin{eqnarray}
  S^{(1)} &=& \sum_{\bra \ssx \ssy \ket}\la_{11}^{(1)}
  \chi_{1\ssx}\chi_{1\ssy} \; , \label{char17}  \\
  S^{(2)}&=& \sum_{\bra \ssx\ssy \ket}\la_{22}^{(2)}\chi_{2\ssx}\chi_{2\ssy}
  +2d\sum_{\ssx}\la_{20}^{(2)} \chi_{2\ssx}  \; , \label{char18} \\
  S^{(3)}&=& \sum_{\bra \ssx \ssy \ket}\Big[
  \la_{11}^{(3)}\chi_{1\ssx} \chi_{1\ssy} +
  \la_{33}^{(3)}\chi_{3\ssx}\chi_{3\ssy}\nonumber \\
  &&\quad\quad
  + \la_{13}^{(3)}
  \left(\chi_{1\ssx}\chi_{3\ssy}+\chi_{3\ssx}\chi_{1\ssy}\right)
  \Big]
  \; .
\label{char19}
\end{eqnarray}
The new couplings $\la_{pq}$ are combinations of the $\kappa$'s,
namely
\begin{eqnarray}
\la_{11}^{(1)}&=&\kappa_1 \; , \qquad
\la_{11}^{(3)}=-\kappa_1 \kappa_2 + \frac{4}{3}\kappa_1^3 \; , \label{char21}\\
\la_{22}^{(2)} &=& \kappa_2 - \ha \kappa_1^2 \; , \qquad
\la_{02}^{(2)}=- \ha\kappa_1^2 \; , \label{char22}\\
\la_{33}^{(3)} &=& \kappa_3 - \kappa_1 \kappa_2 + \frac{1}{3}
\kappa_1^3 \; , \label{char23}\\
\la_{13}^{(3)} &=& - \kappa_1 \kappa_2 + \frac{2}{3} \kappa_1^3 \;
. \label{char24}
\end{eqnarray}
In the asymptotic regime, $\beta_t \ll 1$, the $\la$'s may be
expanded with the help of \refs{char9} (see Section~\ref{sec5}
below). The results (\ref{char17}-\ref{char19}) from combining
character and strong coupling expansion suggest the following
ansatz for the PLA,
\be \label{char27}
  \Spl = \sum_{\left<\ssx\ssy\right>,(pq)}
  \la_{pq} \chi_{p\ssx}\chi_{q\ssy}\equiv \sum_{(pq)} S_{pq} \; .
\ee
The couplings $\la_{pq}$
are symmetric with respect to their indices  $pq$, and $p-q$ is even.
The ansatz \refs{char27} coincides with the one suggested by
Dumitru et al.~\cite{dumitru:03} which was entirely based on
center symmetry. It is obvious that the action \refs{char27} is
center--symmetric as the $\chi_p$ are even/odd functions for $p$
even/odd, $\chi_p (-L) = (-1)^p \chi_p (L)$.

The terms which product $pq = 0$ are localized at single sites and
correspond to `potentials'. Hence the action splits into hopping
terms $T$ and potential terms $V$ in accordance with (\ref{in5}),
\be\label{char29}
  \Spl = \sum_{(pq) \atop pq \neq 0} S_{pq} + 2\sum_{p} S_{p0}
  \equiv T + V   \; .
\ee
With $\chi_0=1$ the potential $V$ has the explicit form
\be\label{char31}
  V = 2d\sum_{p \; \mathrm{even}}\la_{0p}
  \sum_{\ssx} \chi_{p\ssx} \;
  .
\ee
For what follows we need some notation. We will, of course,
truncate our actions at some maximum `spin' $p$, say at $p=r$.
Thus we define the truncated actions
\be \label{SR}
  S_r \equiv \sum_{(pq)}^r S_{pq} = \sum_{(pq) \atop pq \neq 0}^r S_{pq}
  + 2\sum_{p=2}^r S_{p0} \equiv T_r + V_r \;  ,
\ee
where all summations are cut off at $p=r$. Explicitly, the first
few terms are
\begin{eqnarray}
  S_1 &=& S_{11} \; , \label{S1}\\
  S_2 &=& (S_{11} + S_{22}) + 2S_{20} \equiv T_2 + V_2 \; , \label{S2} \\
  S_3 &=& (T_2 + S_{31} + S_{33}) + 2S_{20} \equiv T_3
  + V_3 \; . \label{S3}
\end{eqnarray}
Note that actually $V_3 = V_2$. In the strong--coupling expansion
the action $S_r$ is of order $O(\beta^{rN_t})$. As before, we
refer to $p=1$ (fundamental representation) as the LO, to $p=2$
(adjoint representation) as NLO and so on. In this paper we will
not go beyond $p=3$, a truncation that neglects terms which are
NNNLO in the strong--coupling expansion.

According to (\ref{S2}) the first potential term arising in the
strong--coupling character expansion is of NLO ($\beta^{2N_t}$) and
quadratic in the Polyakov loop $L$,
\be\label{char33}
 V_2 = 2d \la_{02} \sum_{\ssx} \chi_{2\ssx} \; .
\ee

\section{Mean Field Approximation}\label{sec4}
Before we actually relate the PLAs \refs{char27} to Yang--Mills theory
let us analyze their critical behavior which is interesting in
itself. Both via mean--field (MF) analysis and MC simulation we
will see that the models typically have a second order phase
transition at certain critical couplings. Obviously, this should
match with the Yang--Mills critical behavior.

The models involving more than a single hopping term also show a
first order transition at which the order parameter $\langle
L\rangle$ jumps. For the $SU(2)$ case discussed here this
transition is not related to Yang--Mills. Matching to the latter
hence implies that the effective couplings should stay away from
the first--order critical surface.

To develop a MF approximation for the Polyakov loop $L$ with its
nontrivial target space $[-1, 1]$ we use a variational approach
based on the text \cite{roepstorff:90}. Our starting point is the
\emph{effective action} $\Gamma$ for the Polyakov loop dynamics.
This time, the term `effective action' refers to the generating
functional for the 1PI correlators of the Polyakov loop. The
former represents the complete information of the quantum field
theories based on the PLAs $\Spl$. This is obvious from the fact
that $\Gamma$ is the Legendre transform of the Schwinger
functional $W$,
\begin{eqnarray}
  \Gamma[\bar L]&=&(\cL W)[\bar L] \; , \quad
  W[J]=\log Z[J] \; , \label{mf3} \\
  Z[J]&=&\int \cD L \,\exp\Big(- \Spl [L]+(J,L)\Big) \; . \label{mf4}
\end{eqnarray}
These are the standard definitions to be found in any text book on
quantum field theory. For an alternative, variational
characterization of $\Gamma$ \cite{roepstorff:90} we consider the
following \emph{probability measures} on field space,
\be\label{mf7}
  d\mu[L] = \cD L \, p[L] \; ,
\ee
with nonnegative function $p[L]$ and $\cD L$ as in (\ref{imc11}).
Averages are calculated with $\mu$. For example, the mean PLA is
\be\label{mf9}
  \langle \Spl \rangle_\mu = \int d\mu[L] \, \Spl[L] \; ,
\ee
while the Boltzmann--Gibbs--Shannon entropy is given by the
average of $\log p$,
\be \label{mf11}
  S_{\mathrm{BGS}} [\mu] \equiv - \bra \log p \ket_\mu
= -\int d\mu[L] \,  \log p[L] \; .
\ee
The relevant variational principles are obtained as follows. By
subtracting (\ref{mf11}) from (\ref{mf9}) one forms
\be \label{mf12}
  F[\mu] \equiv \bra \Spl + \log p \ket_\mu \equiv F[p] \; .
\ee
This analog of the free energy is varied with respect to $p$
under appropriately chosen
constraints. These are added via Lagrange multipliers. If one just
requires normalization to unity, $\bra 1 \ket_\mu = 1$, one finds
that the probability $p$ for which (\ref{mf12}) becomes extremal
is given by the standard measure, $p = \exp(-\Spl)/Z[0]$.
Inserting this into (\ref{mf12}) yields the infimum
\be
  \inf_\mu F[\mu] = - \log Z[0] \equiv - W[0] \; .
\ee
Comparing with (\ref{mf3}) this may be interpreted as the
effective action for $J = \bar{L} = 0$. If we vary $F$ in
(\ref{mf12}) keeping the expectation value of $L$ fixed at $\bar
L_{\ssx}$ by means of a Lagrange multiplier $J_{\ssx}$, we find
the probability
\be \label{mf12a}
  p[L] = \frac{e^{-S + (J,L)}}{Z[J]} \; ,
\ee
where $J$ is to be viewed as a function of $\bar{L}$, obtained
via inverting the implicit relation $\bar{L} = \delta W / \delta
J$. Plugging (\ref{mf12a}) into (\ref{mf12}) yields a new
variational infimum which is precisely the effective action,
\be \label{mf13}
  \inf_\mu \left\{F[\mu] \left| \bra \Lx \ket_\mu = \bar{L}_{\ssx} \right.
  \right\} = (J, \bar{L}) - W[J] \equiv \Gamma [\bar{L}] \; .
\ee
In the MF approximation for the effective action one minimizes
only with respect to all \emph{product measures} $\nu \in
\mathcal{P}$,
\be \label{mf15}
  d\nu[L] = \prod_{\ssx} d\nu_x(\Lx) \; , \quad
  d\nu_{\ssx}(u) = p_{\ssx}(u) \, dh(u) \; ,
\ee
where $h$ is the reduced Haar measure from (\ref{imc11}). Hence,
the exact expression (\ref{mf13}) is replaced by its MF version
according to
\be\label{mf17}
  \inf_{\nu \in {\cal P}} \left( F[\nu] \, \big \vert \,
  \langle \Lx \rangle_\nu = \bar L_{\ssx}\right) \equiv U_{\rm MF}[\bar L] \; .
\ee
Clearly, from the variational principle, the MF effective action
$U_{\rm MF}$ bounds the effective action $\Gamma$ from above.
Since the set of all product measures is not convex (unlike the
set of all probability measures), the MF action need not be
convex. We may, however, use its convex hull given by the double
Legendre transformation, $\Gamma_{\mathrm{MF}} \equiv \cL^2
\left(U_{\mathrm MF}\right)$, which for non--convex
$U_{\mathrm{MF}}$ will be a better approximation, $\Gamma \le
\Gamma_{\mathrm{MF}}\le U_{\mathrm{MF}}$.

For a product measure the entropy and mean action entering $F$ in
\refs{mf17} turn into sums (of products) of single site
expectation values. With the abbreviation
\be\label{mf21}
  \int d\nu_{\ssx}(u)\, f(u)\equiv\langle f\rangle_{\ssx} \; ,
\ee
we find for the general class of character actions
\refs{char27}
\be \label{F_NU}
  F[\nu] =
  \sum_{\langle \ssvc{x}\ssvc{y} \rangle, (pq)} \hskip-2mm \la_{pq}
  \langle\chi_p\rangle_{\ssx}\langle\chi_q\rangle_{\ssy}
 + \sum_{\ssx}\langle \log p_{\ssx}\rangle_{\ssx} \; .
\ee
This becomes extremal for the single site measure
\be \label{mf27}
  p_{\ssx}(u) = \frac{\exp \big( -V(u) +
  J_{\ssx}u \big)}{z_0(J_{\ssx})}  \; ,
\ee
which replaces (\ref{mf12a}). $V$ denotes the potential from
(\ref{char31}), and the normalization factor is the single--site
partition function,
\be \label{mf29}
  z_0(j) \equiv \int dh(u) \, \exp \big( -V(u) + j u \big) \equiv
  \exp \big( w_0 (j) \big) \; .
\ee
In analogy with (\ref{mf12a}) the Lagrange multiplier (or external
source) $J_{\ssx}$ in the single--site measure $p_{\ssx}$ has to
be eliminated.  This is done by inverting the relation
\be \label{mf31}
  \bar{L}_{\ssx} = w_0'(J_{\ssx}) \; ,
\ee
(the prime henceforth denoting derivatives with respect to $J$) so
that $J{\ssx} = J_{\ssx} (\bar{L}_{\ssx})$. Since the Schwinger
function $w_0$ is strictly convex, the relation between the mean
field $\bar L$ and the external source $J$ in \refs{mf31} is
one-to-one. This will become important in a moment.

Taking these considerations into account the infimum of
(\ref{F_NU}) becomes the MF potential
\be \label{mf33}
  U_{\rm MF}[\bar L] =  \sum_{\langle\ssx\ssy\rangle \atop (pq) \neq 0}
  \la_{pq}\langle\chi_p\rangle_{\ssx}\langle\chi_q\rangle_{\ssy}
  + \sum_{\ssx}\ga_0(\bar L_{\ssx}),
\ee
where all single--site expectation values are calculated with the
measure \refs{mf27} subject to the condition (\ref{mf31}). The
quantity $\ga_0$ is the Legendre-transform of $w_0$,
\be \label{mf35}
  \ga_0(\ell) = \inf_{j} \big( \ell j-w_0(j) \big) \; .
\ee
In this paper we consider \emph{effective potentials} rather than
effective actions and hence assume that the source and mean field
are constant, $\bar L_{\ssx} \equiv \ell$. The effective potential
is just the effective action for constant fields, divided by the
number of lattice points.  The MF potential (\ref{mf33}) then
simplifies to
\be \label{mf37}
  \uMF(\ell) = d \sum_{(pq) \neq 0}
  \la_{pq} \langle \chi_p \rangle \langle \chi_q  \rangle +
  \ga_0(\ell) \; .
\ee
This bounds the true effective potential from above and so does
its convex hull,
\be \label{mf39}
  \ga_{\rm MF}={\cL}^2\big(u_{\rm MF}\big) \; .
\ee
In \cite{fujimoto:88} it is proved that $u_{\rm MF}$ is the MF
approximation to the \emph{constraint effective potential}
\cite{o'raifeartaigh:86}. To calculate $\uMF$ one proceeds as
follows:
\begin{enumerate}
\item
For a given source $j$ one computes the single--site partition
function $z_0$ according to \refs{mf29} and its logarithm, the
Schwinger function, $w_0 = \log z_0$.
\item
The minimizing $j$ in \refs{mf35} satisfies the equation
\be \label{mf41}
  \ell =w_0'(j)=\frac{\int dh(u) \, u \exp\left(ju-V(u)\right)}
  {\int dh(u) \,   \exp\left(ju-V(u)\right)} \; ,
\ee
which must be inverted to yield $j=j(\ell)$.
\item The solution is used to calculate $\ga_0$,
\be
  \ga_0 (\ell) =j(\ell)\ell-w_0(j(\ell)) \; .
\ee
\item Then one computes the expectation values
\be\label{mf43}
  \phantom{xxxx} \langle \chi_p\rangle=\frac{1}{z_0(j)}
\int \exp\big(ju\!-\!V(u)\big)
  \chi_p(u)\,dh(u)\,,
\ee
where $j=j(\ell)$ as obtained from (\ref{mf41}) is inserted
everywhere. Plugging (\ref{mf43}) into \refs{mf37} finally results
in the MF potential $\uMF$.
\end{enumerate}
We are interested in the MF expectation value of the Polyakov loop
which minimizes the MF potential \refs{mf37}. Since the relation
$\ell = \ell(j)$ in \refs{mf41} is one-to-one, the condition that
$\ell$ is a minimum of $\uMF$ is equivalent to $\uMF'=0$. We thus
need the $j$--derivatives of $\ga_0$,
\be\label{mf45}
  \ga_0' \big( \ell(j) \big) = \frac{d\ga_0}{d\ell} \frac{d\ell}{dj}
  = \frac{j}{2} \langle\chi_1\rangle' \; ,
\ee
as well as of the mean characters,
\be\label{mf47}
  2\langle\chi_p\rangle'=
  \langle\chi_{p+1}\rangle+\langle\chi_{p-1}\rangle
  -\langle\chi_p\rangle\langle\chi_1\rangle \; .
\ee
Setting $\uMF' = 0$ results in the \emph{self-consistency
condition}
\be\label{mf49}
  \sum_{(pq) \neq 0}2d\la_{pq}\big(\langle
  \chi_p\rangle\langle\chi_q\rangle\big)'+ j \langle \chi_1 \rangle' =
  0 \; ,
\ee
where all $j$--derivatives are calculated via \refs{mf47}.

Let us finally find the couplings for which the curvature of
$\uMF$ at the origin changes sign. For these, the value of $\uMF$
at $\ell = 0$ turns from a minimum to a local maximum signaling a
second order phase transition. Since the potential in \refs{mf27}
is center symmetric, the strictly convex functions $w_0$ and
$\ga_0$ are both even functions with absolute minimum at the
origin. It follows that $\ell=0$ is mapped to $j=0$ and that (for
these arguments) the second $\ell$--derivative of $\uMF$ is
proportional to its second $j$--derivative,
\be\label{mf51}
  \frac{d^2 \uMF}{d\ell^2} \Big\vert_{\ell=0} = 0 \quad
  \Longleftrightarrow \quad \frac{d^2 \uMF}{d j^2} \Big\vert_{j=0} = 0 \; .
\ee
To calculate the second $j$--derivative of $\uMF$ we need
\be\label{mf53}
  4\langle\chi_p\rangle''\big\vert_{j=0}=
  \big(\langle\chi_{p+2}\!+\!\chi_{p}\!+\!\chi_{p-2} \rangle
  -\langle\chi_p\rangle\langle \chi_2\rangle\big) \big\vert_{j=0}
  \; ,
\ee
which holds for even potentials. Together with \refs{mf47} and
\refs{mf49} this implies the `zero--curvature condition'
\be\label{mf55}
  \sum_{(pq) \neq 0}2d \, \la_{pq}\,
  \big(\langle\chi_p\rangle\langle\chi_q\rangle\big)''\vert_{j=0}+
  \langle\chi_1\rangle'\big\vert_{j=0}=0 \; .
\ee
\subsection{Ising-type models}
If there is no potential in the PLA \refs{char29} and if the
hopping terms contain only NN interactions we refer to the
resulting actions as \emph{Ising-type models}. The associated
actions have been denoted $T_r$ in (\ref{SR}). Setting $V =0$ in
\refs{mf29} we obtain
\be\label{is1}
  z_0(j)=I_0(j)-I_2(j) \; ,
\ee
while the mean characters in \refs{mf43} become
\be\label{is3}
  \langle \chi_p\rangle\equiv \Delta_p(j) = \frac{I_p(j)-I_{p+2}(j)}{I_0(j)
  - I_2(j)} \; .
\ee
For $p=1$ we find the field conjugate to the source $j$,
\be\label{is5}
  \langle \chi_1\rangle=2\ell(j)=
  \frac{I_1(j)-I_{3}(j)}{I_0(j)-I_2(j)} \; .
\ee
The inverse relation $j=j(\ell)$ is used in the MF potential,
\be\label{is7}
  \uMF(\ell)=d\sum_{p q \neq 0}\la_{pq}
  \Delta_p(j)\Delta_q(j)+\ell j-w_0(j) \; ,
\ee
the minimum of which is determined by the self-consistency
condition \refs{mf49}. Since $\Delta_{p\neq 0}(0)$ and
$\Delta_{p\neq 1}'(0)$ both vanish, the condition \refs{mf55} for
a second order phase transition simplifies to
\be\label{is9}
  \la_{11c}=-\frac{1}{2d} \; .
\ee
Hence, in the MF approximation all Ising-type models show a second
order phase transition at the critical coupling $\la_{11c}$. In
the presence of NLO couplings the possibility of first--order
phase transitions arises. In this case, the critical couplings
have to be determined numerically (see below).

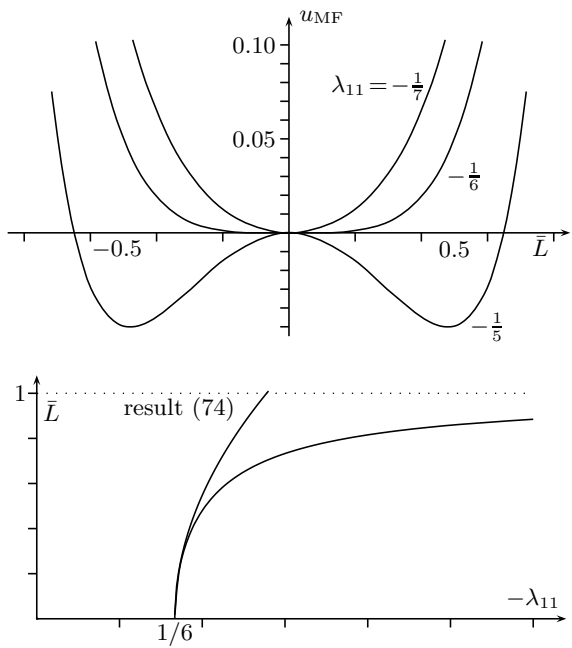
\begin{figure}[ht]
\psset{xunit=4.4mm,yunit=2.5mm,linewidth=0.6pt}
\begin{pspicture}(-9,-6)(8,13)
\psaxes[labels=none,tickstyle=bottom,Dx=2,Dy=1]{->}(0,0)(-8.5,-5.5)(8.5,11.5)
\rput(7.6,-.7){$\bar L$}
\rput(1,11.5){$\uMF$}
\pscurve
(-7.17,7.52)(-6.82,3.08)(-6.39,-0.68)(-5.84,-3.45)
(-5.14,-4.87)(-4.72,-5.00)(-4.25,-4.74)(-3.14,-3.26)
(-1.83,-1.27)(-0.37,-0.06)(0.37,-0.06)(1.83,-1.27)
(3.14,-3.26)(4.25,-4.74)(4.72,-5.00)(5.14,-4.87)
(5.84,-3.45)(6.39,-0.68)(6.82,3.08)(7.17,7.52)
\pscurve
(-5.84,10.19)(-5.14,5.69)(-4.25,2.48)(-3.14,0.70)
(-1.83,0.08)(-0.37,0.00)(0.37,0.00)(1.83,0.08)
(3.14,0.70)(4.25,2.48)(5.14,5.69)(5.84,10.19)
\pscurve
(-4.72,10.27)(-4.25,7.64)(-3.14,3.52)(-1.83,1.04)
(-0.37,0.04)(0.37,0.04)(1.83,1.04)(3.14,3.52)
(4.25,7.64)(4.72,10.27)
\rput[l](-1.7,5){$0.05$}
\rput[l](-1.7,10){$0.10$}
\rput(5,-.9){$0.5$}
\rput(-5.2,-.9){$-0.5$}
\rput(6,-5){$-\frac{1}{5}$}
\rput(5.3,3){$-\frac{1}{6}$}
\rput(2.7,7.8){$\la_{11}\!=\!-\frac{1}{7}$}
\end{pspicture}
\psset{xunit=1.1mm,yunit=.3mm,linewidth=0.6pt}
\begin{pspicture}(-5,-10)(62,120)
\psaxes[labels=none,tickstyle=bottom,Dx=10,Dy=20]{<->}(64,108)
\psline[linestyle=dotted](0,100)(60,100)
\pscurve(16.67,0.02)(17.35,24.02)(18.71,39.37)
(20.07,48.33)(21.43,54.61)(22.79,59.37)(24.15,63.11)
(25.51,66.17)(26.87,68.73)(28.23,70.89)(29.59,72.77)
(30.95,74.39)(32.31,75.82)(33.67,77.10)(35.03,78.24)
(37.76,80.20)(39.12,81.04)(40.48,81.82)(41.84,82.52)
(43.88,83.50)(47.28,84.88)(51.16,86.20)(60.00,88.48)
\rput(60,9){$-\la_{11}$}
\rput(1.8,92){$\bar L$}
\rput(16.7,-7){$1/6$}
\rput(-2,100){$1$}
\psplot{16.67}{28}{x 6 mul 100 sub sqrt 12.247 mul}
\rput[r](24,93){result \refs{is19}}
\end{pspicture}
\caption{\label{fig:isingMF}MF prediction for the effective
potential and order parameter for the Ising-type PLA $S_1 = T_1$
from \refs{is11}.}
\end{figure}

Historically, the first derivation of an effective action for the
confinement--deconfinement phase transition is due to Pol{\'o}nyi
and Szlach{\'a}nyi \cite{polonyi:82}. They have also utilized the
strong--coupling expansion on a Euclidean lattice and already
found the LO contribution from (\ref{S1}),
\be\label{is11}
  S_1 \equiv T_1 = \sum_{\langle\ssx\ssy\rangle} \la_{11} \chi_{1\ssx}
  \chi_{1\ssy} \; .
\ee
Note that our sign-convention is such that \textit{negative}
$\la_{11}$ corresponds to `ferromagnetism'. The action
(\ref{is11}) entails the MF potential
\be\label{is13}
  \uMF^{(1)}(\ell)=4d\la_{11}\ell^2+\ell j-w_0(j) \; ,
\ee
where one inverts the map $j \mapsto \ell$ in \refs{is5} to obtain
$j(\ell)$. This potential is minimal for $\ell=\bar L$ which
solves
\be\label{is15}
  8d\la_{11}\bar L+j(\bar L)=0 \; ,
\ee
implying the self-consistency condition
\be\label{is17}
  2\bar L=\Delta_1\big(-8d\la_{11}\bar L\big) \; .
\ee
The effective potential and order parameter are plotted in
FIG.~\ref{fig:isingMF}.
Near the critical coupling the order parameter has the typical
square root behavior,
\be\label{is19}
  \bar L\approx\sqrt{\frac{3}{2}}\
  \sqrt{\frac{\la_{11}}{\la_{11c}}-1} \quad\hbox{for}\quad
  \vert\la_{11}\vert\searrow \vert\la_{11c}\vert.
\ee
Apart from the second--order transition at $\la_{11c} =-1/2d$
there is no further phase transition (in the MF approximation) for
the LO ansatz \refs{is11} involving only one coupling.

\begin{figure}[ht]
 \hskip-1cm\epsfig{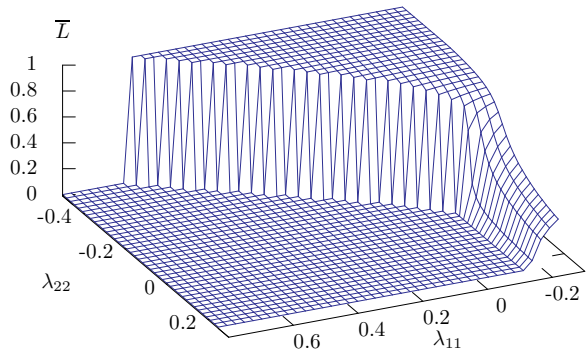}
\caption{\label{fig:isingMF2}MF behavior of the order parameter
$\bar{L}$ for the Ising-type PLA $T_2$ from \refs{is21}.}
\end{figure}

\begin{figure}[ht]
\hskip-1cm\epsfig{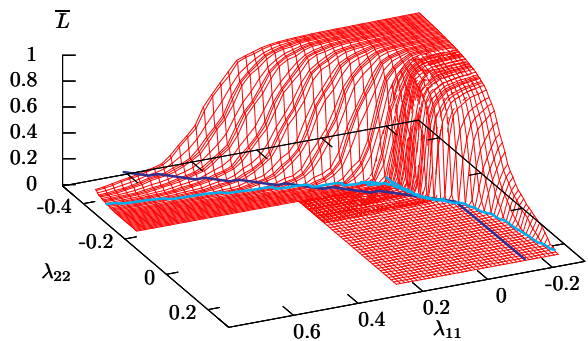}
\caption{\label{fig:isingMF3} MC data for the order parameter
$\bar{L}$ for the Ising-type PLA $T_2$ from \refs{is21}. Shown are
the critical curves of the MF approximation (dark) and MC simulations (bright).
}\end{figure}

As already mentioned, the situation is different for the
Ising--type PLA with NLO coupling $\la_{22}$,
\be\label{is21}
  T_2 = \sum_{\langle\ssx\ssy\rangle}\left(
  \la_{11}\chi_{1\ssx}\chi_{1\ssy} + \la_{22} \chi_{2\ssx}
  \chi_{2\ssy}\right) \; .
\ee
The corresponding MF potential reads
\be\label{is23}
  \tilde{u}_{\mathrm MF}^{(2)}=4d\la_{11} \ell^2 + d \la_{22}
\Delta_2^2(j) + \ell j-w_0(j) \; ,
\ee
where $\Delta_2$ is defined in \refs{is3} and $j(\ell)$ follows
from \refs{is5}. In addition to the second--order phase transition at
$\la_{11c}=-1/6$ the model has a first order transition: the order
parameter jumps as a function of $\lambda_{22}$ as long as
$\la_{11} > \la_{11c}$. The behavior of the order parameter as a
function of the two couplings is shown in FIG.~\ref{fig:isingMF2}.
\\
In FIG.~\ref{fig:isingMF3} the results for the order parameter
$\bar L$ as obtained by MC simulations are plotted. Shown is $\bar
L$ for several thousand sample points in the $(\la_{11},\la_{22})$
plane. Near the second--order phase transition curve with
$\la_{11}\lesssim 0.2$  about $2.5$M updates
were sufficient, whereas near the transition curve with $\la_{11}\gtrsim 0.2$
at least $50$M updates
were required.
\\
We interpret this slow
convergence as a signal for a first--order
phase transition as predicted by the MF approximation,
see FIG.~\ref{fig:isingMF3}.


\subsection{Ginzburg--Landau--type models}
\begin{figure}[ht]
\vskip-8mm \epsfig{file=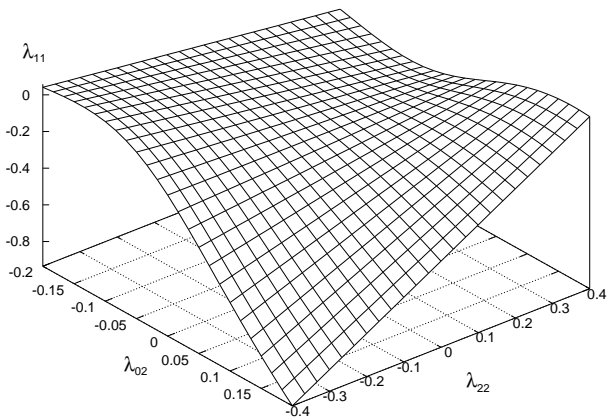,scale=.38,angle=-90}
\caption{\label{fig:lgcs1}Critical surface for the
Ginzburg--Landau PLA $S_2$ from \refs{lg1}.}
\end{figure}

Inspired by the nomenclature in \cite{svetitsky:86} we call
effective actions with a nonvanishing potential as in
\refs{char29} \emph{Ginzburg--Landau--type models}. The
lowest--order potential term is $V_2$ given in \refs{char33} which
is of NLO in the strong--coupling expansion. Hence we first
consider the action
\be\label{lg1}
  S_2 = T_2 + V_2,\quad  V_2 = 2d \la_{02} \sum_{\ssx} \chi_{2\ssx}  \; .
\ee
For $\la_{02} \neq 0$ the MF potentials and mean values must be
computed numerically. Only for the special case of vanishing
source may the relevant integrals be found analytically,
\be\label{lg3}
  \int \!dh(u)\, e^{-2au^2}\chi_{2p}(u)
  = \frac{(-)^p}{e^a}\left[ I_p(a)+I_{p+1}(a) \right] \; .
\ee
For what follows it is useful to define
\begin{eqnarray}
s_{pq}&=&I_p(4d\la_{02})+I_q(4d\la_{02}) \; , \label{lg7} \\
d_{pq}&=&I_p(4d\la_{02})-I_q(4d\la_{02}) \; , \label{lg8}
\end{eqnarray}
so that the mean characters can be written as
\be\label{lg5}
  \langle \chi_{2p}\rangle\big\vert_{j=0}
  =(-)^p\;\frac{s_{p,p+1}}{s_{01}} \; .
\ee
Using (\ref{lg7}) and (\ref{lg8}) the critical surface for
second--order transitions in the $3$-dimensional space of coupling
constants $\la_{11},\la_{22},\la_{02}$ is determined by the
equation
\be\label{lg11}
  \la_{11}=-\frac{1}{2d}\frac{s_{01}}{d_{02}}
  +\la_{22}\left(\frac{s_{12}}{d_{02}}\right)^2
  \left(\frac{s_{03}}{s_{12}}-\frac{s_{12}}{s_{01}}\right) \; ,
\ee
which is plotted in FIG.~\ref{fig:lgcs1}.

The NNLO Ginzburg--Landau--type model is given by
\be\label{lg13}
  S_3 = T_3 + V_3 \; ,
\ee
and has the same potential $V_3 = V_2$ as the NLO action
\refs{lg1} whence \refs{lg5} still holds. It follows that the
critical surface for second--order transitions in the
$5$-dimensional space of coupling constants $\la_{11}, \la_{22},
\la_{02}, \la_{13}, \la_{33}$ is determined by the transcendental
equation
\begin{eqnarray}
  \la_{11}&=&-\frac{1}{2d}\frac{s_{01}}{d_{02}}
  +\la_{22}\left(\frac{s_{12}}{d_{02}}\right)^2
  \left(\frac{s_{03}}{s_{12}}-\frac{s_{12}}{s_{01}}\right)\nn\\
  &&-\la_{33}\left(\frac{d_{12}}{d_{02}}\right)^2
  +2\la_{13}\frac{d_{12}}{d_{02}} \label{lg15} \; .
\end{eqnarray}
For fixed small couplings $\la_{33}$ and $\la_{13}$ this surface
is very close to the one in FIG.~\ref{fig:lgcs1}.

Summarizing we have seen that the effective models undergo both
first-- and second--order phase transitions, and that an MF prediction
for the phase structure of these models agrees well with the
results of MC simulations.

\section{IMC Results for $\boldsymbol{\la_{pq}(\beta)}$}\label{sec5}

Employing our IMC method based on the geometrical Schwinger--Dyson
equations of Section \ref{sec2} we have calculated the couplings
$\la_{pq}$ for different PLAs as functions of $\beta$. For our
lattice with $N_s=20$ and $N_t=4$ the critical Wilson coupling is
$\beta_c=2.30$. Below $\beta_c$ we find the hierarchy
\be\label{nr1}
  |\la_{11}| \gg |\la_{22}| \, , \; |\la_{02}| \gg |\la_{33}| \, , \;
  |\la_{13}| \; ,
\ee
in agreement with the \emph{strong--coupling} expansion
(\ref{char17}-\ref{char19}). According to Ogilvie
\cite{ogilvie:84}, the \emph{weak--coupling} asymptotics of
$\la_{11}$ is linear in $\beta$,
\be \label{nr2}
  \la_{11} (\beta) = - \frac{\beta}{2N_C N_t} + \hbox{const}
\quad (\beta\gg\beta_c) \;, \nn
\ee
which in our case ($N_C = 2$) leads to
\be \label{nr3}
  \la_{11}(\beta)=-\frac{\beta}{16}+\hbox{const}= -0.0625\,\beta +
  \hbox{const} \; .
\ee
We have compared our IMC results for the couplings
$\la_{pq}(\beta)$ with the strong--coupling predictions
(\ref{char7}, \ref{char21}-\ref{char24}) and the weak--coupling
result \refs{nr3}. As expected from our reasoning above, the
lowest order PLAs based on group characters approximate the true
Polyakov--loop dynamics very well in the \emph{strong--coupling}
regime. For weak coupling we find the linear relation \refs{nr3}
already for the LO PLA. For the NLO Ginzburg--Landau--type model
with $3$ couplings the slope $-0.0614$ is very close to the weak
coupling result $-0.0625$ in \refs{nr3}. Thus we are confident
that our PLAs describe the true Polyakov loop dynamics \emph{below
and above} the critical Wilson coupling very well.
\subsection{\label{sec5.1}Leading--order action}
The effective couplings $\la_{11}$ for the LO Ising--type model
\refs{is11} for $\beta$--values below and above the critical
$\beta_c=2.30$ are listed in TABLE~\ref{tablecc1}, Appendix~A. We
read off the value $\la_{11}(\beta_c)=-0.132$. If $S_1$ would be the exact PLA then its
critical coupling $\la_{11c}$ would be $\la_{11}(\beta_c)$. A
direct MC simulation of the action $S_1$ reveals that this model
has critical coupling $\la_{11c}=-0.18$. The MF prediction
$\la_{11c}=-0.17$ comes surprisingly close to the former `true'
value. The critical coupling may alternatively be estimated by
using the strong--coupling results \refs{char9} and \refs{char21}
to calculate $\la_{11}(\beta_c)$ (extrapolating them to $N_t=4$
and $\beta_c=2.30$),
\be \label{lo3}
  \la_{11} = -(\beta/4)^{N_t} \stackrel{\beta=\beta_c}{\approx} -0.11 \; .
\ee
The output of the different methods is compiled in
TABLE~\ref{tablecc0}.
\begin{table}
\caption{\label{tablecc0} Comparison of the critical coupling
values for the LO Ising--type model $S_1$. The `exact' value is
obtained via MC simulation.}
\begin{ruledtabular}
\renewcommand{\arraystretch}{1.3}
\begin{tabular}{l@{\extracolsep\fill}r}
method&critical coupling\\ \hline
MC simulation&$\la_{11c}=-0.18$ \\
MF & $\la_{11c}=-0.17$\\
strong coupling & $\la_{11c}=-0.11$ \\
IMC& $\la_{11}(\beta_c)=-0.14 $\\
\end{tabular}
\end{ruledtabular}
\end{table}
The values obtained are quite close to each other. The `true'
value $-0.18$ stems from simulating \refs{is11} with the MF
approximation coming closest. Somewhat surprisingly, the
strong--coupling result yields the right order of magnitude. The
discrepancy between direct simulation and IMC just means that the
action \refs{is11} represents an oversimplification and does not
match with Yang--Mills well enough. The IMC value
$\la_{11}(\beta_c)=-0.14$ constitutes a compromise equivalent to a
one--parameter fit of the effective to the Yang--Mills
Schwinger--Dyson equations or $n$--point functions.

\begin{figure}[ht]
\psset{xunit=2.9cm,yunit=27cm,linewidth=0.4pt,dotsize=1.6mm,dotsep=2pt}
\begin{pspicture}(1,-0.32)(3.6,0.01)
\psaxes[Ox=1,Dx=.5,Dy=0.1,Oy=-.3,axesstyle=frame](1,-.3)(3.6,0.005)
\psplot[linestyle=dashed]{1}{2.3}{x 4 div 4 exp neg}
\rput(1.13,-0.02){$\la_{11}$}
\rput(3.3,-0.29){$\beta$}
\rput(2.3,-0.31){$\beta_c$}
\psline[linestyle=dotted](1,-0.1667)(3.6,-0.1667)
\rput(1.2,-0.155){MF}
\psline[linestyle=dotted](2.3,-.3)(2.3,0.005)
\rput(3.15,-0.04){$S_1$}
\psset{linecolor=red}
\psline[linestyle=dashed](2.4,-0.17155)(3.5,-0.21469)
\psdots[dotstyle=o]
(2.95,-0.041)
(1.10,-0.0047)(1.70,-0.0305)(2.20,-0.104)
(2.28,-0.1231)(2.29,-0.1267)(2.30,-0.1325)
(2.32,-0.1512)(2.34,-0.1585)(2.38,-0.1658)(2.60,-0.1792)
(2.80,-0.1874)(3.00,-0.1952)(3.50,-0.2146)
\psset{linecolor=magenta}
\rput(3.15,-0.08){$S_2$}
\psline[linestyle=dashed](2.4,-0.19983)(3.5,-0.26736)
\psdots[dotstyle=asterisk]
(2.95,-0.082)
(1.10,-0.0045)(1.70,-0.0299)(2.20,-0.1165)
(2.28,-0.1441)(2.29,-0.1485)(2.30,-0.1563)
(2.32,-0.1786)(2.34,-0.1873)(2.38,-0.1957)(2.60,-0.2125)
(2.80,-0.2242)(3.00,-0.2362)(3.50,-0.2676)
\psset{linecolor=blue}
\rput(3.15,-0.06){$T_2$}
\psline[linestyle=dashed](2.4,-0.18955)(3.5,-0.24513)
\psdots[dotstyle=x]
(2.95,-0.062)
(1.10,-0.0046)(1.70,-0.0303)
(2.20,-0.1033)(2.28,-0.1234)(2.29,-0.1276)(2.30,-0.1349)
(2.32,-0.1598)(2.34,-0.1697)(2.38,-0.1799)(2.60,-0.1989)
(2.80,-0.2102)(3.00,-0.2206)(3.50,-0.2447)
\psset{linecolor=gray}
\psline[linestyle=dashed](2.4,-0.20936)(3.5,-0.28314)
\rput(3.15,-0.1){$S_3$}
\psdots[dotstyle=triangle]
(2.95,-.102)
(1.1,-0.004)(1.7,-0.029)(2.2,-0.116)(2.28,-0.143)
(2.29,-0.148)(2.30,-0.156)(2.32,-0.179)(2.34,-0.1892)
(2.38,-0.1999)(2.6,-0.2227)(2.8,-0.236)(3,-0.250)(3.5,-0.283)
\end{pspicture}
\caption{\label{fig:la11}\textsl{ The coupling $\la_{11}$ as a
function of $\beta$. Dashed curve for $\beta<\beta_c$: asymptotic
behavior \refs{lo3}; dashed lines for $\beta>\beta_c$: linear fits
$\la = a \beta +b$; data points: IMC
results from Tables~\ref{tablecc1}-\ref{tablecc3} in Appendix~A.}}
\end{figure}
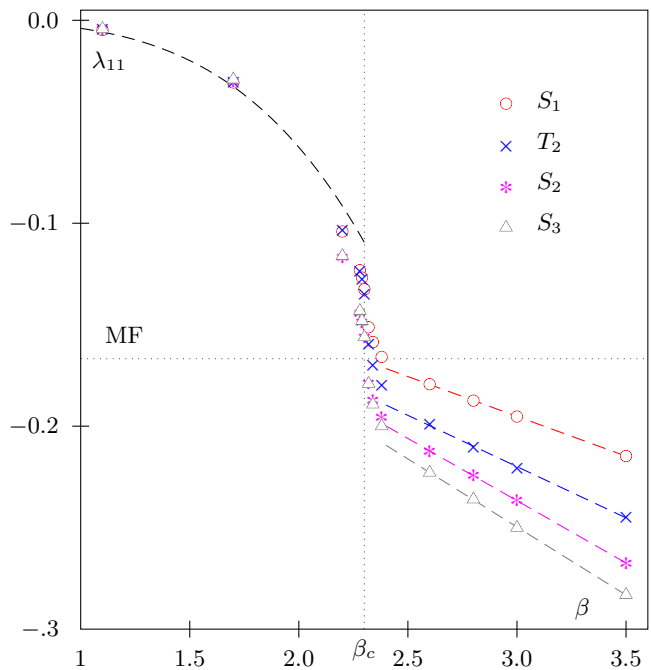

In FIG.~\ref{fig:la11} we compare the results for $\la_{11}$ from
strong-- and weak--coupling expansion with the data of our IMC
simulations. Note that the asymptotic formula \refs{lo3} works up
to $\beta \simeq 2$. The IMC values for the LO action $S_1$ in
\refs{is11} are marked with circles.
Note further that the dependence $\la_{11}(\beta)$ is indeed
linear in the weak--coupling regime, $\la_{11}=-0.0392\,\beta
-0.0774$, in accordance with the prediction \refs{nr3} for the
weak--coupling asymptotics. The slope, however, turns out being too
small. This will be remedied in what follows by including more
couplings.


\subsection{Next--to--leading--order actions}

The effective action \refs{is11} has been confirmed and extended
by several authors \cite{ogilvie:84,drouffe:84,green:84,gross:84}.
We have also checked its generalizations by considering the NLO
Ising--type model \refs{is21} without potential terms and the
Ginzburg--Landau action \refs{lg1} containing all NLO terms from
\refs{S2}.

The NLO IMC results for $\la_{11}(\beta)$ are displayed in
FIG.~\ref{fig:la11} (crosses and asterisks), those for
$\la_{22}(\beta)$ in FIG.~\ref{fig:la22} (same symbols) and those
for $\la_{02}(\beta)$ in FIG.~\ref{fig:la02} (asterisks). The
error bars for the couplings are listed in the tables in
Appendix~A. Upon comparing the values for $\la_{11}$ in
FIG.~\ref{fig:la11} (or TABLE~\ref{tablecc1}) we note that for
$\beta < \beta_c$ the predictions for $\la_{11}$ are almost model
independent. In the weak--coupling regime, on the other hand, they
are less stable. Hence, adding terms to the PLA may change the
coupling constants considerably in this regime.

For the NLO actions the dependence $\la_{11}(\beta)$ is linear in
the weak coupling regime, similarly as for the LO action $S_1$.
However, the slopes $a$ in the linear fits $\la_{11}(\beta)=a
\beta + b$ above $\beta_c$ are model dependent. For the LO and NLO
PLA they are given in TABLE~\ref{tableslopes} together with the
weak--coupling result \refs{nr3}.
\begin{table}
\caption{\label{tableslopes} Slopes of the linear fits to the
weak--coupling asymptotics (\protect\ref{nr2}) of $\la_{11}$.}
\begin{ruledtabular}
\renewcommand{\arraystretch}{1.3}
\begin{tabular}{c@{\extracolsep\fill}cccc}
model&$1/2N_C N_t$ & $S_1$
& $T_2$ & $S_2$ \\ \hline
slope&-0.0625 &-0.0392 &-0.0505 &-0.0614\\
\end{tabular}
\end{ruledtabular}
\end{table}
For the Ising--type models without potential terms the slope is
not reproduced very well, and we conclude that in the
weak--coupling or high--temperature regime potential terms should
be included for an accurate description of the Polyakov--loop
dynamics. Indeed, the slope for the action \refs{lg1} with
potential term $V_2$ is almost identical to the prediction
\refs{nr2} of the weak--coupling asymptotics.

FIG.~\ref{fig:la22} shows the dependence of the adjoint coupling
constant $\la_{22}$ on the Wilson coupling $\beta$. Also shown is
the prediction \refs{char22} of the strong coupling expansion using
\refs{char9},
\be \label{nlo9}
  \la_{22}(\beta)=-\frac{113}{162}\left({\beta\over
4}\right)^8\approx -0.69753\left({\beta\over 4}\right)^8 \; , \ee
which again reproduces the IMC results for $\beta< 2$.

\begin{figure}[ht]
\psset{xunit=2.8cm,yunit=145cm,linewidth=0.4pt,dotsize=1.6mm,dotsep=2pt}
\begin{pspicture}(.8,-.012)(3.6,0.05)
\psaxes[Dx=0.5,Dy=0.01,Ox=1,Oy=-.01,axesstyle=frame](1,-.01)(3.6,0.05)
\rput(1.14,0.047){$\la_{22}$}
\rput(3.3,-.0075){$\beta$}
\rput(2.3,-.012){$\beta_c$}
\psplot[linestyle=dashed]{1}{2.2}{0.69753 x 4 div 8 exp mul neg}
\psline[linestyle=dotted](2.3,-.01)(2.3,0.05)
\rput(1.4,0.04){$T_2$}
\psset{linecolor=blue}
\psdots[dotstyle=x]
(1.2,0.0398)
(1.10,-0.001)(1.70,-0.001)(2.20,-0.0020)(2.28,0)(2.29,0.002)
(2.30,0.006)(2.32,0.023)(2.34,0.03)(2.38,0.037)(2.60,0.0382)
(2.80,0.035)(3.00,0.0319)(3.50,0.0264)
\rput(1.4,0.035){$S_2$}
\psset{linecolor=red}
\psdots[dotstyle=asterisk]
(1.2,0.0348)
(1.10,0)(1.70,-0.001)(2.20,0)(2.28,0.0029)(2.29,0.004)(2.30,0.006)
(2.32,0.017)(2.34,0.022)(2.38,0.026)(2.60,0.0260)
(2.80,0.0229)(3.00,0.0193)(3.50,0.0107)
\rput(1.4,0.03){$S_3$}
\psset{linecolor=gray}
\psdots[dotstyle=triangle]
(1.2,0.0298)
(1.10,0)(1.70,0)(2.20,0)(2.28,0.002)(2.29,0.003)(2.30,0.006)
(2.32,0.019)(2.34,0.026)(2.38,0.035)(2.60,0.04)(2.80,0.035)
(3.00,0.031)(3.50,0.02)
\end{pspicture}
\caption{\label{fig:la22}\textsl{ The adjoint coupling $\la_{22}$
as a function of $\beta$ for the NLO and NNLO actions. Dashed
curve: asymptotic behavior \refs{nlo9}; data points: IMC results
from Tables~\ref{tablecc1}-\ref{tablecc3}.}}
\end{figure}
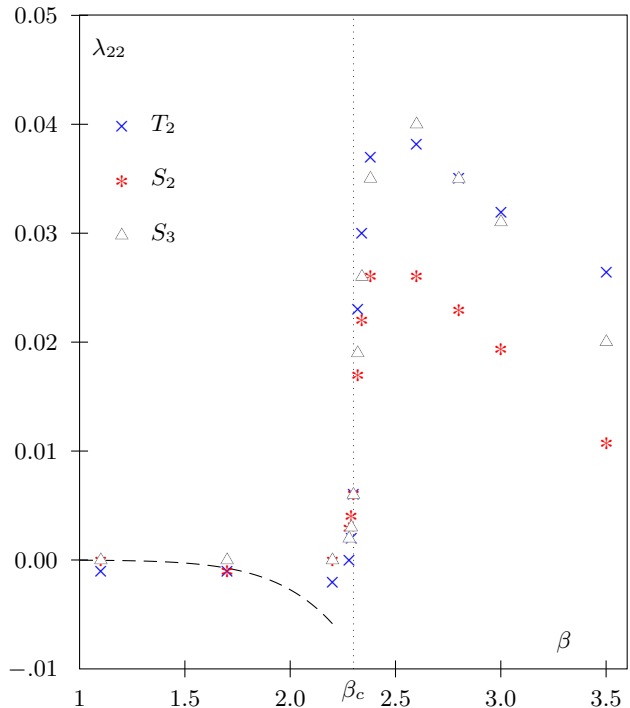\noindent
The critical coupling
\be
\la_{22}(\beta_c)=-0.008[1]
\ee
is the same for both PLAs \refs{is11} and \refs{lg1}. Up to
$\beta_c$ the two actions have almost identical couplings
$\la_{22}$. It seems that for $\beta>\beta_c$ these couplings also
depend linearly on $\beta$, as was the case for $\la_{11}(\beta)$.

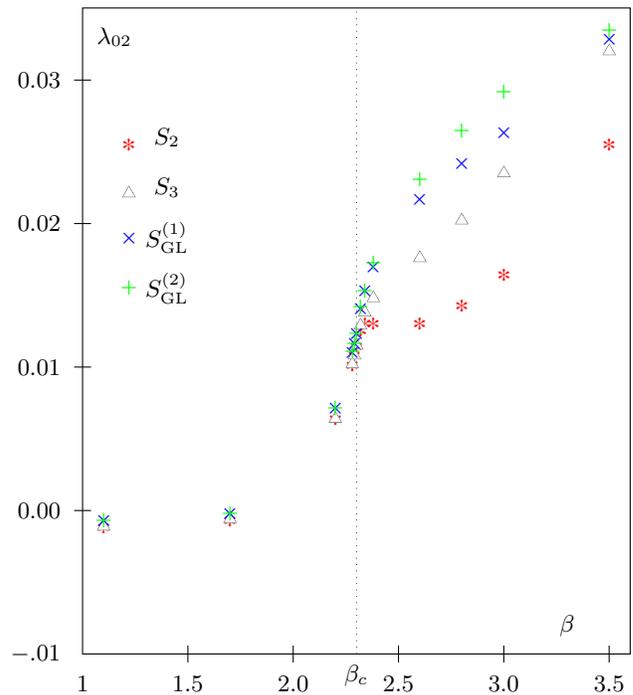
\begin{figure}[ht]
\psset{xunit=2.8cm,yunit=191cm,linewidth=0.4pt,dotsize=1.6mm,dotsep=2pt}
\begin{pspicture}(.8,-.012)(3.6,0.035)
\psaxes[Dx=0.5,Dy=-0.01,Ox=1,Oy=-.01,axesstyle=frame](1,-0.01)(3.6,0.035)
\rput(1.15,0.033){$\la_{02}$}
\rput(3.3,-.008){$\beta$}
\rput(2.3,-.0115){$\beta_c$}
\psline[linestyle=dotted](2.3,-.01)(2.3,0.035)
\rput(1.4,0.026){$S_2$}
\psset{linecolor=red}
\psdots[dotstyle=asterisk]
(1.22,0.0255)
(1.1,-0.0012)(1.7,-0.0007)(2.2,0.0064)
(2.28,0.0101)(2.29,0.0108)(2.3,0.0114)(2.32,0.0125)
(2.34,0.013)(2.38,0.013)(2.6,0.013)(2.8,0.0143)
(3,0.0164)(3.5,0.0255)
\rput(1.4,0.0225){$S_3$}
\psset{linecolor=gray}
\psdots[dotstyle=triangle]
(1.22,0.0221)
(1.1,-0.0011)(1.7,-0.0006)(2.2,0.0064)
(2.28,0.0102)(2.29,0.0108)(2.3,0.0115)(2.32,0.0129)
(2.34,0.0138)(2.38,0.0148)(2.6,0.0176)(2.8,0.0202)
(3,0.0235)(3.5,0.032)
\rput(1.4,0.019){$S_{\rm GL}^{(1)}$}
\psdots[linecolor=blue,dotstyle=x]
(1.22,0.019)
(1.1,-0.00066)(1.7,-0.00019)(2.2,0.00712)
(2.28,0.01103)(2.29,0.01164)(2.3,0.0123)
(2.32,0.01405)(2.34,0.01527)(2.38,0.01697)
(2.6,0.02172)(2.8,0.02416)(3,0.02633)
(3.5,0.03281)
\rput(1.4,0.0155){$S_{\rm GL}^{(2)}$}
\psdots[linecolor=green,dotstyle=+]
(1.22,0.0155)
(1.1,-0.00069)(1.7,-0.00018)(2.2,0.00715)
(2.28,0.01106)(2.29,0.01164)(2.3,0.01238)
(2.32,0.0142)(2.34,0.01531)(2.38,0.01723)
(2.6,0.02305)(2.8,0.0265)(3,0.02914)
(3.5,0.03348)
\end{pspicture}
\caption{\label{fig:la02}\textsl{ The coupling $\la_{02}$ as a
function of the Wilson coupling $\beta$ for NLO, NNLO and
Ginzburg--Landau actions; data points: IMC results from
Tables~\ref{tablecc2}-\ref{tablecc4}.}}
\end{figure}

\subsection{Next-to-next-to-leading order}

We have seen that the NLO approximations describe the Polyakov
loop dynamics very well in the symmetric strong--coupling phase.
In the weak--coupling regime, however, there is still room for
improvement. Hence we have calculated the five coupling constants
$\la_{11},\la_{22},\la_{02},\la_{13},\la_{33}$ appearing in the
general NNLO PLA $S_3$ for several values of the Wilson coupling.
This action is the sum of all terms up to order $\beta_t^{\,3N_t}$
in the strong--coupling expansion. As expected, adding the third
order terms $S_{31}$ and $S_{33}$ does not change the lower--order
couplings $\la_{11},\la_{22},\la_{02}$ (as obtained via $S_2$) in
the broken phase. This can be seen in  FIG.s~\ref{fig:la11},
\ref{fig:la22} and \ref{fig:la02}, where the IMC results for the
NNLO action $S_3$ from \refs{S3} are depicted by triangles. The
numerical values for these couplings and the couplings $\la_{33}$
and $\la_{13}$ together with their statistical errors are given in
Appendix~A.

\subsection{\label{sec5d} Ginzburg--Landau models}

In his review \cite{svetitsky:86}, Svetitsky has suggested to
emphasize the potential term in \refs{in5} by specializing to an
ansatz of Ginzburg--Landau type,
\be \label{gl1}
  S_{\mathrm{GL}} =S_1 + \sum_{\ssx} V(\Lx) \; ,
\ee
with center--symmetric potential \refs{char31}. Replacing the
coefficients $\la_{0p}$ of the characters in this potential by
\begin{eqnarray}
  \la_4&=&2d\cdot 4^2\la_{04} \; , \\
  \la_2&=&2d\cdot 4(\la_{02}-3\la_{04}) \; , \label{gl3}
\end{eqnarray}
leads for $p\leq 4$ to the even polynomial
\be \label{gl5}
  V(L)=\la_2 L^2+\la_4 L^4+\rm{const} \; .
\ee
In Table~\ref{tablecc4} we have listed the couplings for the
models with quadratic and quartic potentials,
\begin{eqnarray}
  S^{(1)}_{\mathrm{GL}}&=& S_1
  + \la_2\sum_{\ssx}\Lx^2\label{gl7} \; , \\
  S^{(2)}_{\mathrm{GL}}&=& S_{\mathrm{GL}}^{(1)}
  + \la_4\sum_{\ssx}\Lx^4\label{gl9} \; ,
\end{eqnarray}
obtained via IMC within our standard range of $\beta$. The values
for $\la_{11}$ both for strong and weak coupling are almost
identical to those of the NLO model $S_2$. For this reason we have
refrained from plotting them in FIG.~\ref{fig:la11}. The potential
couplings are important in the broken weak--coupling phase where
they become sizable. The coupling $\la_{02}$ for the
Ginzburg--Landau models (\ref{gl7},\ref{gl9}) is shown in
FIG.~\ref{fig:la02}.

\subsection{Summary}

The values for the couplings of the different PLAs arising at
critical Wilson coupling $\beta_c=2.30$ are listed in
TABLE~\ref{tableccc}. They are almost model independent.
$\la_{11}$, in particular, is always close to the MF value
$-0.167$ (for the Ising type models).
\begin{table}
\caption{\label{tableccc} Effective couplings of different PLAs at
$\beta = \beta_c = 2.30$.}
\begin{ruledtabular}
\renewcommand{\arraystretch}{1.4}
\begin{tabular}{c@{\extracolsep\fill}rrrrr}
model&$S_1$&$T_2$&
$S_2$&$S_3$&$S_{\mathrm{GL}}^{(2)}$\\ \hline
$\la_{11}(\beta_c)$&$-0.133$&$-0.135$&$-0.156$&$-0.156$&$-0.155$\\
$\la_{22}(\beta_c)$&&$0.006$&$0.006$&$0.006$&\\
$\la_{02}(\beta_c)$&&&0.011&$0.012$&$0.012$\\
\end{tabular}
\end{ruledtabular}
\end{table}

The couplings below and above $\beta_c$ and their statistical
errors are compiled in Appendix~A. There one may also find
$\la_{33},\la_{13}$ and $\la_{04} = \la_4 / 32d$ for different
Wilson couplings.

The stability of the couplings for $\beta < \beta_c$ is a strong
indication that (in this regime) the Yang--Mills ensemble is very
well approximated already by the NLO models with $2$ or $3$
couplings. The results of the following section will further
confirm this statement.

The Ising--type coupling $\la_{11}$ becomes a linear function of
$\beta$ in the weak--coupling regime, in accordance with the
weak--coupling prediction \refs{nr2}. For the NLO action the slope
is $-0.0614$ which compares favorably with the weak--coupling
value $-0.0625$. For the Ginzburg--Landau--type actions the slope
is almost identical to the one of the NLO models.

The Ising--type couplings change rapidly at the critical Wilson
coupling $\beta_c=2.30$ as demonstrated in FIG.s~\ref{fig:la11}
and \ref{fig:la22}. For example, the coupling $\la_{11}$ decreases
from $-0.1$ below $\beta_c$ to $-0.2$ above $\beta_c$. This jump
of $\la_{11}$ forces the system into the ferromagnetic phase. For
$\la_{22}$ the jump is even more dramatic, from $0$ to $0.04$. The
potential couplings $\la_{02}$ and $\la_{04}$ change more smoothly
when the systems changes from the symmetric to the broken phase.


\section{Two--point functions}
\begin{figure}[ht]
\psset{xunit=3.7mm,yunit=26cm,linewidth=0.6pt,dotsize=1.6mm}
\begin{pspicture}(-0.1,-0.04)(22,0.28)
\hskip8mm
\psaxes[Dx=2,Oy=-.01,Dy=0.05,axesstyle=frame](0,-.01)(20,0.275)
\rput(10,0.25){$\beta=2.20$}
\rput[l](0.3,0.265){$G_{x0}$}
\rput(19.4,0.003){$x$}
\psline[linestyle=dotted,dotsep=3pt](0,0)(20,0)
\rput[l](12,0.201){YM}
\psdots[dotstyle=+]
(11,0.20)
(0,0.25036)(1,0.03482)(2,0.00897)(3,0.00316)(4,0.00131)
(5,0.00059)(6,0.00026)(7,0.00013)(8,0.00003)
(9,-0.00003)(10,-0.00004)(11,-0.00003)(12,0.00003)
(13,0.00013)(14,0.00026)(15,0.00059)(16,0.00131)
(17,0.00316)(18,0.00897)(19,0.03482)(20,0.25036)
\rput[l](12,0.161){$S_2$}
\psdots[linecolor=red,dotstyle=asterisk]
(11,0.16)
(0,0.25116)(1,0.03090)(2,0.00406)(3,0.00054)(4,0.00007)
(5,0)(6,-0.00001)(7,0)(8,0)(9,0)(10,0)(11,0)(12,0)(13,0)
(14,-0.00001)(15,0)(16,0.00007)(17,0.00054)(18,0.00406)
(19,0.03090)(20,0.25116)(20,0.25116)
\rput[l](12,0.181){$T_2$}
\psdots[linecolor=blue,dotstyle=x]
(11,0.18)
(0,0.2589)(1,0.02885)(2,0.00339)(3,0.00043)(4,0.00006)(5,-0.00001)
(6,-0.00002)(7,-0.00001)(8,-0.00001)(9,-0.00002)(10,-0.00002)
(11,-0.00002)(12,-0.00001)(13,-0.00001)(14,-0.00002)(15,-0.00001)
(16,0.00006)(17,0.00043)(18,0.00339)(19,0.02885)(20,0.2589)(20,0.2589)
\end{pspicture}
\caption{\label{twopt_SY_2.2} \textsl{The Yang--Mills
(\textrm{YM}) two--point function compared to the ones obtained
from the NLO effective actions $T_2$ and $S_2$;
$(N_s,N_t)=(20,4)$.}}
\end{figure}
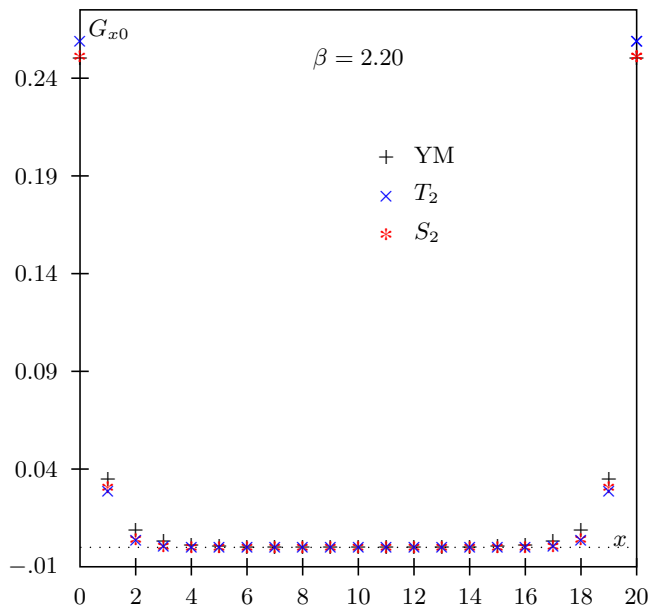

\begin{figure}[ht]
\psset{xunit=3.7mm,yunit=20cm,linewidth=0.6pt,dotsize=1.6mm}
\begin{pspicture}(-0.1,-.03)(22,0.41)
\hskip8mm
\psaxes[Ox=0,Dx=2,Dy=0.05,axesstyle=frame](0,0)(20,0.4)
\rput(8,0.37){$\beta=3.0$}
\rput[l](12,0.36){YM}
\rput[l](12,0.335){$T_2$}
\rput[l](12,0.31){$S_2$}
\rput[l](12,0.285){$S_3$}
\rput[l](0.3,0.38){$G_{x0}$}
\rput(19.4,0.01){$x$}
\psdots[dotstyle=+]
(11,0.36)
(0,0.3527)(1,0.23772)(2,0.22628)(3,0.22435)(4,0.22389)
(5,0.22372)(6,0.22369)(7,0.22366)(8,0.22367)(9,0.22364)
(10,0.22363)(11,0.22364)(12,0.22367)(13,0.22366)(14,0.22369)
(15,0.22372)(16,0.22389)(17,0.22435)(18,0.22628)(19,0.23772)
(20,0.3527)(20,0.3527)

\psdots[linecolor=red,dotstyle=asterisk]
(11,0.31)
(0,0.34327)(1,0.20612)(2,0.18312)(3,0.1784)(4,0.17728)(5,0.17698)
(6,0.17689)(7,0.17687)(8,0.17684)(9,0.17683)(10,0.17684)(11,0.17683)
(12,0.17684)(13,0.17687)(14,0.17689)(15,0.17698)(16,0.17728)
(17,0.1784)(18,0.18312)(19,0.20612)(20,0.34327)(20,0.34327)
\psdots[linecolor=blue,dotstyle=x]
(11,0.335)
(0,0.34531)(1,0.1927)(2,0.16666)(3,0.16101)(4,0.15957)(5,0.15916)
(6,0.15903)(7,0.15897)(8,0.15896)(9,0.15895)(10,0.15895)(11,0.15895)
(12,0.15896)(13,0.15897)(14,0.15903)(15,0.15916)(16,0.15957)
(17,0.16101)(18,0.16666)(19,0.1927)(20,0.34531)(20,0.34531)
\psdots[dotstyle=triangle]
(11,0.285)
(0,0.34634)(1,0.21786)(2,0.19808)(3,0.19435)(4,0.19353)
(5,0.19336)(6,0.19332)(7,0.19332)(8,0.19331)(9,0.19331)
(10,0.19333)(11,0.19331)(12,0.19331)(13,0.19332)(14,0.19332)
(15,0.19336)(16,0.19353)(17,0.19435)(18,0.19808)(19,0.21786)
(20,0.34634)(20,0.34634)
\end{pspicture}
\caption{\label{twopt_SY_3.0} \textsl{The Yang--Mills
(\textrm{YM}) two--point function compared to the one obtained
from the NLO ($T_2$ and $S_2$) and NNLO ($S_3$) effective actions;
$(N_s,N_t)=(20,4)$.}}
\end{figure}
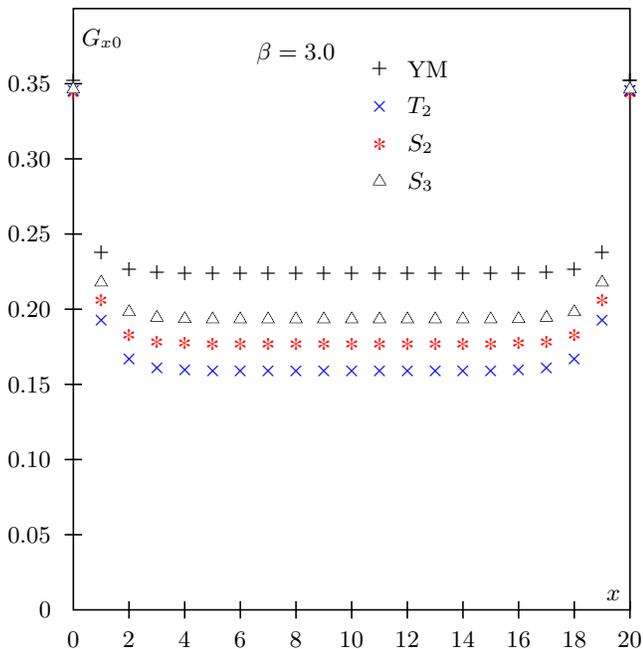

Let us finally check the quality of our PLAs which, after all,
should represent approximations to Yang--Mills theory. To this end
we compare the Yang--Mills two--point function at different
$\beta$--values with those of our effective models inserting the
couplings $\la_{pq}(\beta)$ obtained via IMC. For the NLO actions
at $\beta=2.2$ these couplings are displayed in
TABLE~\ref{table22},
\begin{table}
\caption{\label{table22} Effective couplings for NLO actions at
$\beta = 2.2$.}
\renewcommand{\arraystretch}{1.3}
\begin{ruledtabular}
\begin{tabular}{c@{\extracolsep\fill}ccc}
$\beta=2.2$&$\la_{11}$&$\la_{22}$&$\la_{20}$\\ \hline
$T_2$&0.1033&0.0019&\\
$S_2$&0.1168&0.00042&0.0064\\
\end{tabular}
\end{ruledtabular}
\end{table}
while for $\beta = 3.0$ we have the NLO and NNLO couplings of
TABLE~\ref{table30}.
\begin{table}
\caption{\label{table30} Effective couplings for NLO and NNLO
actions at $\beta = 3.0$.}
\begin{ruledtabular}
\renewcommand{\arraystretch}{1.3}
\begin{tabular}{c@{\extracolsep\fill}ccccc}
$\beta=3$&$\la_{11}$&$\la_{22}$&$\la_{20}$&$\la_{33}$&$\la_{31}$\\ \hline
$T_2$&0.2207&0.0320&&&\\
$S_2$&0.2361&0.0194&0.0165&&\\
$S_3$&0.2506&0.0311&0.0236&0.0035&0.0040\\
\end{tabular}
\end{ruledtabular}
\end{table}
With these couplings we have simulated the models with actions
\be
  T_2 \, , \quad S_2 \, , \quad S_3 \, , \quad S_{\rm GL}^{(1)} \, , \quad S_{\rm
  GL}^{(2)} \, ,
\ee
and calculated the two--point functions displayed in
FIG.s~\ref{twopt_SY_2.2} and \ref{twopt_SY_3.0}. As expected, the
agreement in the center--symmetric phase ($\beta =2.2$) is very
good, while deep in the broken phase ($\beta = 3.0$) there appears
to be room for improvement.
For $\beta=2.2$ the two--point functions of the three effective
models considered are almost identical to the Yang--Mills
two--point function. The data points for $S_3$ and $S^{(2)}_{\rm
GL}$ cannot be distinguished from those for the NLO model $S_2$
and hence are not displayed in FIG.~\ref{twopt_SY_2.2}.

For $\beta=3.0$ the two--point functions and the expectation value
of the mean field are model dependent. In FIG.~\ref{twopt_SY_3.0}
we have plotted the two--point function of Yang--Mills theory and
of the NLO and NNLO effective actions.

For the NLO approximation $T_2$ the value for the condensate is
approximately 20 percent below the Yang--Mills value. Including
potential terms ($S_2$) and NNLO terms ($S_3$) improves the
approximation somewhat as FIG.~\ref{twopt_SY_3.0} shows. The
two--point function for the Ginzburg--Landau action $S^{(2)}_{\rm
GL}$ is almost identical to the one of $T_2$ (and thus has not
been displayed). This implies that also higher--order Ising (or
hopping) terms are important when $\beta > \beta_c$.

\section{\label{sec:7}Discussion}

To really obtain satisfactory approximations to Yang--Mills
expectation values (e.g.\ two--point functions) for all
$\beta$--values one has to go beyond nearest--neighbor
interactions in the effective theory. This has been done in
\cite{dittmann:03} where operators with $r \equiv |\vc{x} -
\vc{y}|$ up to $\sqrt{2}$ (`plaquette operators') were included.
We briefly recapitulate the results of this brute--force
calculation. By reformulating the Schwinger--Dyson equations
\refs{imc15} in terms of characters, the couplings have been
determined by our standard inverse Monte--Carlo routines. We found
that the couplings decrease rapidly not only if we go to higher
representations (i.e.~larger $p$) as above, but also if we
increase the number of links within the plaquette operators used.
The leading $r=1$ term $S_1$ of \refs{S1} with coupling $\la_{11}$
dominates by one order of magnitude compared to the terms with $r
= \sqrt{2}$. This clearly indicates that the effective
interactions are short--ranged in accordance with the
Svetitsky--Yaffe conjecture. Simulating the effective action
including plaquette operators with the couplings obtained via IMC
we have calculated the two--point function in both phases.
Including a total of $14$ couplings the matching between
Yang--Mills and the effective action becomes perfect both in the
broken and symmetric phase \cite{dittmann:03}. It should, however,
be stressed that this brute--force numerical calculation does not
provide too much of physical intuition. We believe that the
present paper, in particular the mean--field analysis, improves
upon \cite{dittmann:03} in this respect.

The most straightforward generalization of our analysis obviously
is to go to $SU(3)$ Yang--Mills theory where one expects a first
order phase transition. Work in this direction is under way.
\begin{appendix}
\section{Tables}
The following TABLEs \ref{tablecc1}--\ref{tablecc4} contain the
effective couplings $\la_{pq}$ with statistical errors for various
values of $\beta$.
\begin{table}[ht]
\caption{IMC results for the effective coupling $\la_{11}$ in the LO action
$S_1$ and the two couplings $\la_{11},\la_{22}$ in $T_2$.}
\label{tablecc1}
\noindent
\begin{tabular*}{8.5cm}{c|@{\extracolsep\fill}r|rr}\hline \hline
Yang-Mills-$\beta$&$S_1: \la_{11}$&$T_2: \la_{11}$&$T_2: \la_{22}$\\
\hline
1.70 &  -0.0305[4] &-0.030[2]&-0.001[1]\\
2.20 &  -0.1040[2]&-0.103[1]&-0.0019[7]\\
2.28 & -0.1231[3] &-0.123[1]&0.000[1]\\
2.29 & -0.1267[4] &-0.127[1]&0.002[1]\\
2.30 & -0.1325[5]&0.135[1]&0.006[1]\\
2.32 & -0.1512[3]&-0.1598[9]&0.023[1]\\
2.34 & -0.1585[2] &-0.1697[6]&0.030[1]\\
2.38 & -0.1658[1] &-0.1799[4]&0.037[1]\\
2.60 & -0.1792[1] &-0.1989[4]&0.0382[8]\\
2.80 & -0.1874[1] &-0.2103[5]&0.0350[8]\\
3.00 & -0.1952[1] &-0.2206[6]&0.0319[8]\\
3.50 & -0.2146[1] &-0.2446[6]&0.0264[6]\\
4.00 & -0.2343[1] &-0.2670[9]&0.0225[6]\\ \hline\hline
\end{tabular*}
\end{table}
\begin{table}[ht]
\caption {IMC results for the three couplings
in $S_2$.}
\label{tablecc2}
\begin{tabular*}{8.4cm}{c|@{\extracolsep\fill}rrr}\hline\hline
$\beta$ & $\la_{11}$ & $\la_{22}$ & $\la_{20}$\\
\hline
1.10 & -0.004[4] & -0.000[2] & -0.0012[5]\\
1.70 & -0.030[4] & -0.001[2] & -0.0007[5]\\
2.20 & -0.116[2] & 0.000[2]  & 0.0064[5] \\
2.28 & -0.144[2] & 0.002[2]  & 0.0101[5] \\
2.29 & -0.148[2] & 0.004[2]  & 0.0108[5] \\
2.30 & -0.156[2] & 0.006[2] & 0.0114[5]  \\
2.32 & -0.178[2] & 0.017[2] & 0.0125[6]  \\
2.34 & -0.187[1] & 0.022[2] & 0.0130[6]  \\
2.38 & -0.195[1] & 0.026[2] & 0.0130[6] \\
2.60 & -0.212[1] & 0.026[2] & 0.013[1] \\
2.80 & -0.224[1] & 0.022[2] & 0.014[1]  \\
3.00 & -0.236[1] & 0.019[2] & 0.016[1]  \\
3.50 & -0.267[1] & 0.010[2] & 0.025[1] \\
4.00 & -0.299[1] & 0.003[2] & 0.037[1]\\ \hline\hline
\end{tabular*}
\end{table}
\begin{table}[ht]
\caption {IMC results for five couplings in $S_3$.}
\label{tablecc3}
\begin{tabular*}{8.4cm}{c|@{\extracolsep\fill}rrrrr}\hline\hline
$\beta$ & $\la_{11}$ & $\la_{22}$ & $\la_{20}$ & $\la_{33}$ & $\la_{31}$\\
\hline
1.10 &-0.004[1] & 0.000[1] & -0.0011[2] & 0.000[1] & 0.0003[6] \\
1.70 &-0.029[2] & 0.000[1] & -0.0006[2] & 0.0001[8] & 0.0000[6] \\
2.20 &-0.116[1] & 0.000[1] & 0.0064[2] & 0.0015[7] & 0.0000[6] \\
2.28 &-0.143[1] & 0.002[1] & 0.0102[3] & 0.0015[8] & 0.0002[6] \\
2.29 &-0.148[1] & 0.003[1] & 0.0108[3] & 0.0015[8] & 0.0000[6] \\
2.30 &-0.156[1] & 0.006[1] & 0.0115[3] & 0.001[1] & 0.0000[6] \\
2.32 &-0.179[1] & 0.019[1] & 0.0129[3] & -0.0006[9] & -0.0004[5] \\
2.34 &-0.1892[8] & 0.026[1] & 0.0138[3] & -0.0014[8] & -0.0014[6] \\
2.38 &-0.1999[6] & 0.035[1] & 0.0148[4] & -0.0032[9] & -0.0026[5] \\
2.60 &-0.2227[8] & 0.040[1] & 0.0176[5] & -0.005[1] & -0.0041[4] \\
2.80 &-0.236[1] & 0.035[1] & 0.0202[8] & -0.0041[9] & -0.0040[4] \\
3.00 &-0.250[1] & 0.031[1] & 0.0235[8] & -0.0034[8] & -0.0039[4] \\
3.50 &-0.283[1] & 0.020[1] & 0.032[1] & -0.002[1] & -0.0032[6] \\
4.00 &-0.291[3] & 0.003[1] & 0.029[2] & -0.0033[9] & 0.0029[8] \\
\hline\hline
\end{tabular*}
\end{table}
\begin{table}[ht]
\caption{IMC results for the couplings $\la_{11},\la_2,\la_4$
in $S^{(2)}_{\rm GL}$.}\label{tablecc4}
\begin{tabular*}{8.4cm}{c|@{\extracolsep\fill}rrr}\hline\hline
$\beta $& $\la_{11}$&$\la_2$&$\la_4$\\ \hline
2.20&-0.110 & 0.186 & -0.002\\
2.25&-0.119 & 0.237 & -0.003\\
2.28&-0.127 & 0.288 & -0.006\\
2.29&-0.128 & 0.303 & -0.007\\
2.30&-0.157 & 0.453 & -0.020\\
2.32&-0.173 & 0.621 & -0.057\\
2.34&-0.176 & 0.698 & -0.093\\
2.40&-0.190 & 0.979 & -0.256\\
\hline\hline
\end{tabular*}
\end{table}
\section{Numerical Details}
All MC calculations have been performed on a $20^3 \times 4$
lattice for which the critical Wilson coupling is $\beta_c=2.30$.
The simulations have been done for $\beta$ ranging from 1.1 to
4.0. We have used a standard `pseudo-heatbath' algorithm
\cite{kennedy:85, fabricius:84} due to Miller \cite{miller:01}.

The IMC routine has been implemented as follows. For
each action term $S_{pq}$ and site we have chosen
the operator
\begin{eqnarray}
  \label{nd:3}
  G_{pq,\ssz} \equiv \frac{1}{\lambda_{pq}} \fud{S_{pq}}{\Lz} \; ,
\end{eqnarray}
which leads to the Schwinger--Dyson equations
\begin{equation}
  \label{eq:sdeqn}
  \sum_{(pq)} \lambda_{pq}\bra(1\!-\!\Lx^2)G_{pq,\ssz} S_{pq,\ssx} \rangle =
  \bra (1\!-\!\Lx^2) G_{,\ssx}-3\Lx G \ket \; .
\end{equation}
Due to translational invariance the coefficients of $G_{\ssz}$ and
$G_{\ssz'}$ are equal if $|\vc{x}-\vc{z}|=|\vc{x}-\vc{z}'|$. In
order to have a sufficiently overdetermined system (for fixed
$pq$) we choose the $N_s$ operators $G_{pq,d}$, $d = 0\ldots N_s$.
Independent of our choice of PLA we have always used the full set
of operators up to truncation values $p,q = 5$, i.e.
\begin{equation}
  \label{eq:4}
  G_{11},G_{22},G_{20},G_{33},G_{31},G_{44},G_{42},G_{40},G_{55}
\end{equation}
with $0 \le d \le 8$ leading to a total of 81 operators (and
equations). Translational invariance admits to use
the spatial average of each Schwinger--Dyson equation
and every configuration. The overdetermined system
is then solved via least--square methods.
We have checked that the couplings obtained in this
way follow a normal distribution, as expected.
Hence we calculated the standard
deviation $\sigma$ and took $2\sigma$ as our error.
Autocorrelation effects have been eliminated via binning. Our
statistics (5k to 10k configurations) entail a statistical error
of $10^{-4}$ which translates into an uncertainty for the
couplings in the NLO action of the order of a few percent.
The NNLO couplings $\lambda_{33}$ and $\lambda_{31}$,
however, have statistical errors of about $20\%$.

Systematic errors are mainly due to the dependence of the
couplings on the operator bases used in the Schwinger--Dyson
equations.

\end{appendix}

\begin{acknowledgments}
The authors thank L.~Dittmann for his collaboration at
an early stage of this work and A.~Dumitru for
useful discussions. T.H. is indebted to
A.~Hart, R.~Pisarski and A.~Khvedelidze for sharing valuable
insights. T.K. acknowledges useful hints from
A.~Sternbeck and D.~Peschka.
\end{acknowledgments}


\end{document}